\documentclass[aps,prl,twocolumn,showpacs,superscriptaddress]{revtex4}

\usepackage{graphicx}

\usepackage[dvips, colorlinks=true, citecolor=blue, urlcolor=blue, linkcolor=blue, ]{hyperref}

\usepackage{amssymb,amsmath}

\usepackage{bbm}

\begin{document}

\title{Aggregation and fragmentation dynamics of inertial  particles in chaotic flows}

\author{Jens C. Zahnow}

\affiliation{Theoretical Physics/Complex Systems, ICBM, University of Oldenburg, 26129 Oldenburg, Germany}

\author{Rafael D. Vilela}

%\email{rdvilela@pks.mpg.de} 

\affiliation{Max Planck Institute for the Physics of Complex Systems, 01187 Dresden, Germany}

\author{Ulrike Feudel}
\affiliation{Theoretical Physics/Complex Systems, ICBM, University of Oldenburg, 26129 Oldenburg, Germany}

\author{Tam\'as T\'el}

\affiliation{Institute for Theoretical Physics, E\"otv\"os University - P.O. Box 32, H-1518, Budapest, Hungary}


%\thanks{}%

%\subjclass{}%

%\keywords{}%

%\date{}%

%\dedicatory{}%

%\commby{}%

% ----------------------------------------------------------------

\begin{abstract}
Inertial particles advected in chaotic flows often accumulate in strange attractors. While moving in these fractal sets they usually approach each other and collide. Here we consider inertial particles aggregating upon collision. The new particles formed in this process are larger and follow the equation of motion with a new parameter.
These particles can in turn fragment when they reach a certain size or shear forces become sufficiently large.
The resulting system consists of a large set of coexisting dynamical
systems with a varying number of particles. We find that the combination of aggregation and fragmentation leads to an asymptotic steady state. The asymptotic particle size distribution depends on the mechanism of fragmentation.
The size distributions resulting from this model are consistent with
those found in rain drop statistics and in stirring tank experiments. 

% For fragmentation due to shear the shape of this size distribution is self-preserving with respect to aggregate strength.
%%, whereas the short-time size distribution of the particles is not.
\end{abstract}

\maketitle

% ----------------------------------------------------------------
There is an increasing recent interest in the advection of inertial particles in fluid flows \cite{fsp_strange_attractors}. This comes in part from the fact that the dynamics of these particles are dissipative, which leads in most flows to a preferential accumulation on chaotic, fractal attractors. Previous studies concentrated mainly on noninteracting particles, in spite of the fact that accumulation leads unavoidable to strong mutual interactions of different kinds. 

Here we consider the interaction of these particles in form of aggregation and fragmentation. When two particles come sufficiently close, they aggregate to a larger one, observing mass and momentum conservation. If the size exceeds a certain threshold value, which depends on a property, the stickiness $\gamma$, of the particles, or other conditions are fulfilled, they break up into smaller pieces. This is the basic mechanism underlying such processes in nature like raindrop formation in clouds \cite{prup} or the sedimentation of marine aggregates \cite{marine_snow} in the ocean. We demonstrate that for the study of such processes a particle based approach may be a useful addition to the usual population-balance equation approach \cite{Thomas}. The latter is based on the assumption of well-mixed particles while our approach takes the incomplete mixing of inertial particles in fluids explicitly into account.

Although concepts of dynamical systems theory can usefully be applied, we show that the entire dynamics is much more complex than that of any usual dynamical system. The dynamics of particles of any size are governed by the same type of equations of motion, but with different parameters since new particles will have new radii. Even if one
  considers a finite number, $n$, of possible sizes (size classes), there are $n$ equations of motion with different size-dependent parameters. We thus have a union of $n$ dynamical systems and, moreover, the number of particles in each size class is changing in time. It is useful to interpret the attractors of the different size classes (of the non-interacting problem) as the skeleton of the full dynamics. Aggregation/fragmentation generates transitions from one attractor to another one. It is this permanent wandering among different attractors which characterizes the new dynamics.  

We show that the combination of aggregation/fragmentation, superimposed on
chaotic inertial advection dynamics, leads to a convergence to an asymptotic
steady state, and this steady state is unique for the cases studied here. We
find that the dynamics and the steady state depend on the fragmentation rule. For fragmentation due to shear we 
present a simple scaling relationship for the asymptotic average size of the
particles. Furthermore, the shape of the asymptotic size distribution can be represented in a scaled form independent of the stickiness $\gamma$.

For simplicity we consider spherical aerosoles, i.e. particles much denser than the ambient fluid and assume that the difference between their velocity $\dot{\bf r}$ and the fluid velocity  ${\bf u}={\bf u}({\bf r}(t),t)$ at the same position is sufficiently small so that the drag force is proportional to this difference (Stokes drag). The dimensionless form of the governing equation for the path ${\bf{r}}(t)$ of such aerosols subjected to  drag and gravity, reads as \cite{maxey_riley}:
\begin{equation}
\ddot{\bf r}=A\left({\bf u}-\dot{\bf r}-W{\bf n}\right), 
\label{maxey}
\end{equation}
where ${\bf n}$ is a unit vector pointing upwards in the vertical direction. Throughout this paper we consider the vertical direction along the axis $y$. The inertia parameter $A$ (larger values for smaller particle size) can be written in terms  of the densities  $\rho_p$ and $\rho_f$ of the aerosol and of the fluid, respectively, the radius $a$ of the aerosols, the fluids kinematic viscosity $\nu$, and the characteristic length $L$ and velocity $U$ of the flow.
 It is $A={R}/{St}$, where $R={\rho_f}/{\rho_p}\ll 1$ is the density ratio and $St=(2a^2U)/(9\nu L)$ is the so-called Stokes number of the aerosol \cite{obs}.
$W=2 a^2 \rho_p g/(9 \nu \rho_f U)$ is the dimensionless settling velocity in a medium at rest. 

Every particle produces perturbations in the flow that decay inversely proportional to the distance from the particle \cite{Happel}. Here we assume a dilute regime, where the local concentration of particles is low enough, so that particle-particle interaction can be neglected \cite{concentration}.

During aggregation and fragmentation the radius of particles changes and so do the parameters $A$ and $W$. The smallest \textit{(primary)} particles considered in this model have dimensionless radius $a_{1}={5}/{30^{1/3}}\times10^{-5}$, mass $m_1=\rho_p {4}/{3}\pi a_1^3$, inertia parameter $A_{1}=7$ and settling velocity $W_{1}=0.4/A_{1}$. All larger particles are assumed to consist of an integer number of these primary particles. $30$ different size classes are considered. A particle that consists of $\alpha$ ($\alpha=1,...,30$ is called \textit{size class index}) primary particles has a radius $a_{\alpha}=\alpha^{1/3}a_{1}$, an inertia parameter $A_{\alpha}= ({a_{1}}/{a_{\alpha}})^2 A_{1}=\alpha^{-2/3} A_1$ and a settling velocity $W_{\alpha}= {\alpha}^{2/3} W_{1}$. The largest particle therefore has a radius $a_{30}=5\times10^{-5}$.

{\em Aggregation} %(in raindrop formation called coagulation) 
takes place upon collision, i.e. if two particles, say of  radius $a_i$ and $a_j$,  come closer than a threshold. Mass conservation requires the radius of the new particle to be $a^3_{new}=a^3_i+a^3_j$. For the size class index this implies a linear rule: $\alpha_{new}=\alpha_i+\alpha_j$ which determines the new inertia parameter via $A_{\alpha_{new}}=\alpha_{new}^{-2/3} A_1$. The velocity of the new particle follows from momentum conservation.

{\em Fragmentation:} we apply one of the following rules. (i) \textit{Size-limiting fragmentation}: If a particle becomes larger than the maximum radius $a_{30}$, it is broken up into two smaller fragments whose radii are chosen randomly, with a uniform distribution between $a_1$ and half the original radius. If any fragment is larger than $a_{30}$ this process is repeated, until no fragment exceeds $a_{30}$.
(ii) \textit{Shear fragmentation} takes place if the velocity gradient is too large. More specifically, the velocity gradient is evaluated
across each particle both in the horizontal and in the vertical direction. If the maximum in any direction exceeds a threshold value, the particle is broken up into two smaller parts in the same way as for size-limiting fragmentation. While size-limiting fragmentation is dominant for raindrops \cite{prup}, shear fragmentation determines the break-up of marine aggregates \cite{Flesch}. 
 
Since for marine aggregates the threshold gradient becomes smaller for larger particles \cite{Flesch}, we write
\begin{equation}\label{condition3}
(\mbox{grad}({\bf u}))_{th}=\gamma {a_{1}}/{a}= \gamma \alpha^{-1/3}.
\end{equation}
Coefficient  $\gamma$ represents the 'stickiness' of the particles. Whatever rule is taken, the result is the reversed process of aggregation: two new particles are formed from an old one with the size class indices: $\alpha_{i,new}+\alpha_{j,new}=\alpha_{old}$. The centers of the new particles are placed along a line segment in a random direction
 so that their distance equals the sum of their radii. Momentum is conserved. For simplicity we assume that the new particles have the same velocity as the old one.
Shear fragmentation is applied together with size-limiting fragmentation to keep the maximum number of occurring size classes at $30$.

At the instant of both the aggregation and the fragmentation process there is a sudden change in the dynamics: the number of particles jumps in $3$ among the $30$ available dynamical systems defined by the size classes. 

For convenience, we treat the case where the fluid flow is two-dimensional, therefore the phase space of the advection dynamics is 4-dimensional. We use the convection model of \cite{Chandrasekhar} with dimensionless velocity field
\begin{equation}
{\bf u}(x,y,t)= [1+k\sin(\omega t)]\left(\begin{array}{c}
\sin(2\pi x)\cos(2\pi y) \\
-\cos(2\pi x)\sin(2\pi y)
\end{array}\right)~,
\end{equation}
where $k=2.72$ is the amplitude and $\omega=\pi$ is the frequency of the periodic forcing. The fluid flow itself is laminar, but the dynamics of the inertial particles can be chaotic. Because of the spatial periodicity of the flow and the resulting spatial periodicity of the attractors the total particle mass, {\bf $M$}, in each $1\times 1$ unit cell remains the same over time. The dynamics can therefore be restricted to one cell. The characteristic size and velocity of the flow are therefore $L=1$, $U=1$, respectively.

%Figure 1: Chaotic attractors in three different size classes: alpha=1,16,30.
\begin{figure}[htb]
		\centering
		\includegraphics[width=0.46\textwidth]{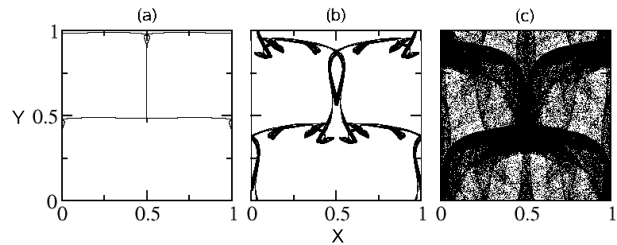}
		\caption{\label{fig:three_attractors} Poincar\'e section of the attractors of
		  Eq. (\ref{maxey}) projected onto the plane of the flow for
		  inertia parameters 
(a) $A=7$ (size class 1), (b) $A=2.778$ (size class 16), (c) $A=2.253$ (size
		  class 30). The positive Lyapunov exponents are (a)
		  $\lambda_1=0.108$, (b) $\lambda_1=0.061$, (c)
		  $\lambda_1=0.119$,  $\lambda_2=0.014$. The settling velocity is $W=0.4/A$.} 	
\end{figure}

In the numerical realization of the problem the particles are advected without
any interaction over a time interval $\delta t=T/20$ at the end of which first
aggregation and then fragmentation take place, instantly. This is repeated
after every time step $\delta t$. To carry out the aggregation process, the
distance between particles is calculated and all particles within a distance less than the sum of their radii aggregate.

 As initial condition we take $10^5$ particles in the smallest size class and no particles in other size classes. Furthermore particles are uniformly distributed over the entire configuration space with velocities matching that of the fluid. This choice fixes the total mass of the system to be $M=10^5 m_1$.

Before presenting the results obtained for the full dynamics, it is instructive to see the attractors of the non-interacting problem. Fig. \ref{fig:three_attractors} presents the attractors for the smallest, an intermediate and the largest size classes. The extension of the attractor seems to grow almost monotonically with the size class index, except for a few intermediate size classes ($\alpha=9...14$), where the attractor size decreases or the attractor becomes periodic.

%Figure 2: Time dependency of the number of particles in three size classes
%and total number of particles and space distribution of particles at short time and close to steady state
\begin{figure}[htb]
		\centering
		\includegraphics[width=0.46\textwidth]{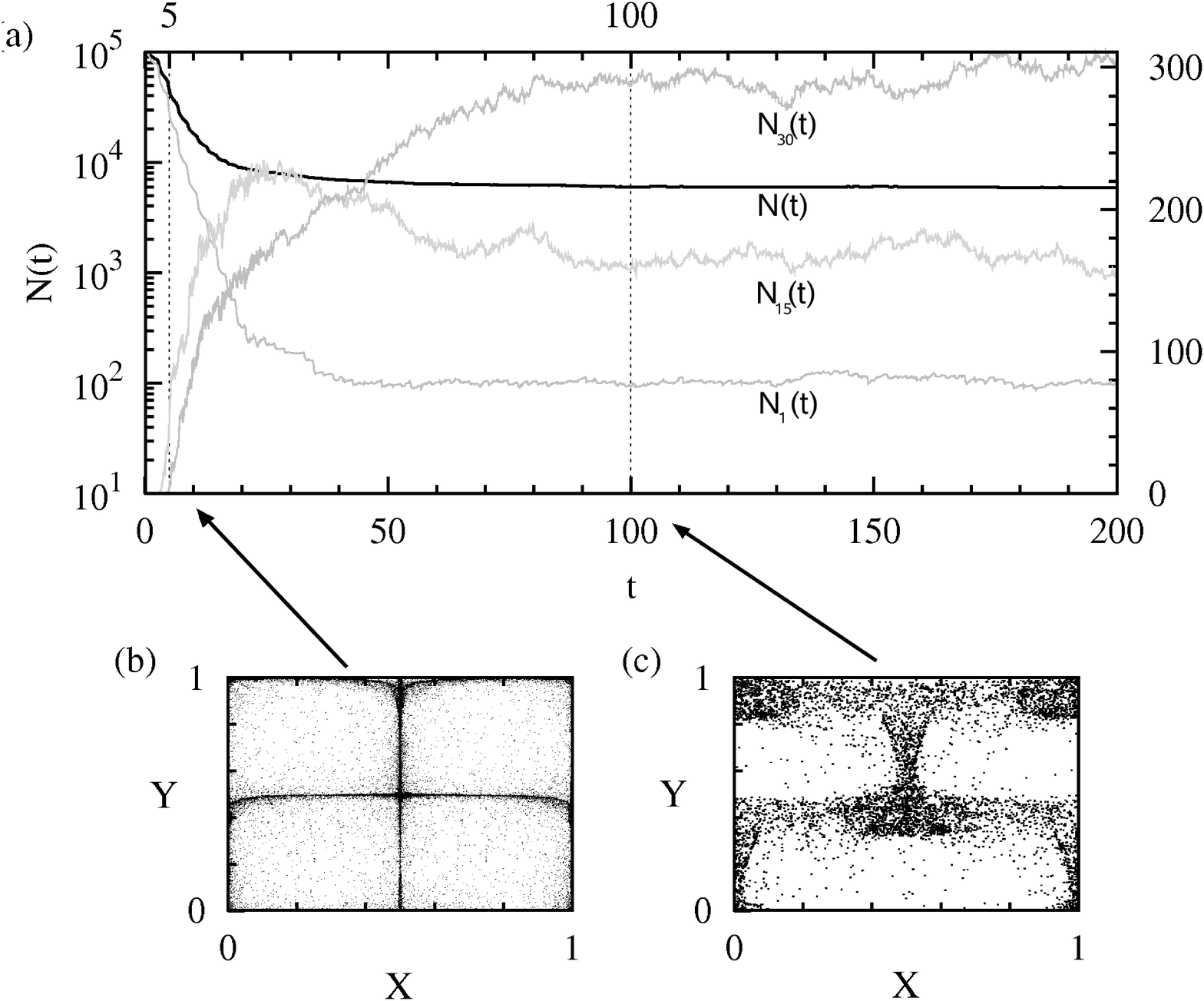}
 		\caption{\label{fig:average_spaceplot} Particle numbers
 		vs. time, and space distributions for size-limiting fragmentation. (a)
 		Total number $N(t)$  of particles (bold - left axis) and the
 		number of particles $N_{\alpha}(t)$ in size class $\alpha$ (gray)
 		for $\alpha=1$ (left axis), $\alpha=16$ (right axis) and
 		$\alpha=30$ (right axis). Distribution of all particles in configuration space at time (b) $t=5$, (c)  $t=100$. }	
\end{figure}

In order to understand the full dynamics, we include first the simplest fragmentation process, the size-limiting fragmentation. Fig. \ref{fig:average_spaceplot}(a) shows the time dependence of the number $N_\alpha(t)$ of particles in a few size classes. The particles leave the initial size class very quickly. After 20 time units nearly all other size classes are considerably occupied. In fact, the population in size class 16 reaches a maximum here, but decreases again later on. It is
the occupation of the largest size classes which continuously increases and then saturates. The total number $N(t)$ of particles (bold line) rapidly decreases first, but saturates later on.  The spatial distribution of particles (Fig. \ref{fig:average_spaceplot}(b/c)) shows that they move initially among the more localized attractors characteristic of small size class indices. Later, the distribution becomes more extended in configuration space when size
classes with extended attractors become well occupied, although the total number of particles is much less than in the initial
phase. While the full dynamics is dominated by transients in-between attractors, the shape of the backbone attractors is clearly recognizable in the plots.

%Figure 3: Average size class and standard deviation (as error bars) over time

\begin{figure}[htb]
		\centering
		\includegraphics[width=0.46\textwidth, height=3.6cm]{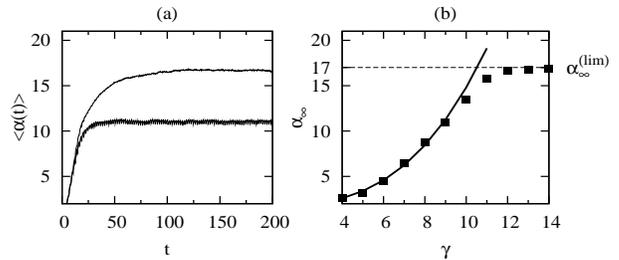}
		\caption{\label{fig:average} (a) Average size class
		    index $\left<\alpha(t)\right>$ vs. time for size-limiting 
		    fragmentation (upper curve) and  shear fragmentation with $\gamma=9$ (lower curve). (b)
Asymptotic average size class index for the same initial particle distribution as a function of the stickiness parameter $\gamma$. Squares: numerical results, continuous  line: fit $\alpha_{\infty}(\gamma)$ by a linear function of $\gamma^{3}$, based on the data for $\gamma<10$}	
\end{figure}

To follow the convergence towards an asymptotic state, we found useful to consider the average size class index $\left<\alpha(t)\right> = \sum_{i=1}^{30}\alpha_{i} N_{\alpha_{i}}(t)/N(t)$.  Fig. \ref{fig:average}(a) shows the time dependence of this index for both types of fragmentation. It illustrates the convergence to an asymptotic steady state for both fragmentation rules. Initially, aggregation leads to a fast increase in the average particle size class 
for both fragmentation rules. Then fragmentation sets in and a balance between aggregation and fragmentation is reached, with a different asymptotic average particle size $\alpha_\infty=\lim\limits_{t\rightarrow\infty}\left<\alpha(t)\right>$ for the two rules. For size-limiting fragmentation the value of $\alpha_{\infty}$ is almost constant over time, while for shear fragmentation $\alpha_{\infty}$ oscillates with the period $T$ of the flow. This is caused by the periodic change in the fluid flow and the corresponding change in the shear forces.

For size-limiting fragmentation, $\alpha_{\infty}$ is, in a broad range, independent of $M$. For shear fragmentation 
with $M<3\times 10^{5} m_1$, $\alpha_{\infty}(M)$ increases approximately linearly with $M$, while for higher values a saturation of $\alpha_{\infty}(M)$ sets in, which is due to size-limiting fragmentation.

By considering other initial conditions than mentioned above, while keeping the total mass $M$ fixed, the asymptotic state is found for both rules to be independent of the chosen initial condition, but for shear fragmentation the asymptotic state does depend on the value of the stickiness $\gamma$. 
 
%Figure 4 Steady state of the size distribution for random maximum and random shear splitting for gamma=9.
\begin{figure}[htb]
		\centering
		\includegraphics{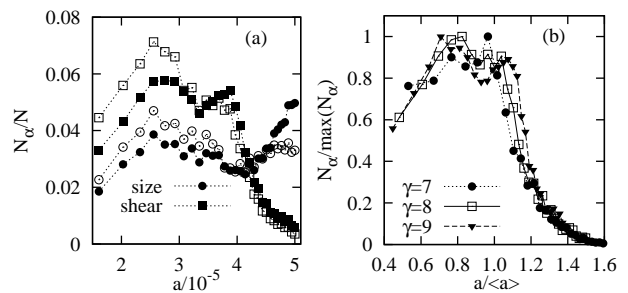}
		\caption{\label{fig:steady_states} Histogram of the particle size distribution. (a) Size class
 percentage $N_{\alpha}/N$ in the steady state versus the dimensionless radius $a$, for size-limiting and shear fragmentation 
with $\gamma=9$. Results for fragmentation into three parts is also shown (open markers). (b) Normalized number of particles versus the relative radius $a/\left<a\right>$ for different values of $\gamma$ (binary splitting).}	

\end{figure}

To illustrate this dependence of the steady state on the stickiness $\gamma$, Fig. \ref{fig:average}(b) shows how $\alpha_\infty$ changes with the stickiness parameter, at a fixed $M$. 
A drastic increase of $\alpha(\gamma)$ can be observed in the interval $4 <\gamma <10$. It is clear that $\alpha_{\infty}$ increases 
with $\gamma$, because particles become more resistant to shear. A quantitative estimate of the shape of this $\alpha_{\infty}(\gamma)$ curve can be derived by assuming that the threshold velocity gradient is approximately constant for the size class index  $\alpha_{\infty}$. From (\ref{condition3}) it then follows that $\alpha_{\infty}$ depends linearly on  $\gamma ^3$. This simple dependence is expected to hold for relatively small values of $\gamma$ and $\alpha_{\infty}$, where shear fragmentation dominates. It can be seen that for higher values of $\gamma$, when size-limiting fragmentation becomes important the $\alpha_{\infty}(\gamma)$ curve deviates from this estimate and converges towards a limiting value $\alpha_{\infty}^{(lim)}$ (Fig. \ref{fig:average}(b)).

% %Figure 5: Short term behavior for t shear splitting for gamma=10!! 
% \begin{figure}[htb]
% 		\centering
% 		\includegraphics[width=0.46\textwidth]{Figure5.eps}
% 		\caption{\label{fig:short_term_dist} Size distributions of the particles in the initial phase ($t=5$) for  size-limiting and shear fragmentation ($\gamma=9$). Solid line: $k_{1}\approx -2.42\times10^5$, dashed line: $k_{2}\approx -2.56\times10^5$.}	
% 
% \end{figure}

In addition to the average quantities it is natural to investigate the occupation of the different size classes in the steady state. Fig. \ref{fig:steady_states}(a) shows the steady state histograms vs. the dimensionless radius. For size-limiting fragmentation the distribution shows one broad peak around smaller size classes and a second, smaller peak at large size classes with a sharp drop-off towards zero beyond the maximum size.
This behavior, with two maxima and a sudden drop after the second peak, is similar to that of observed cloud drop spectra \cite{prup}. For shear fragmentation the steady state distribution also shows two peaks, but much closer together, with a long tail in the particle distribution towards larger sizes that goes smoothly towards zero. For $\gamma<5$ this distribution is not fully developed and only shows one peak. In the intermediate $\gamma$ range, where the distribution is fully developed, but size-limiting fragmentation is not important, a scaling form $\frac{N_{\alpha}}{max(N_{\alpha)} } = f \left( \frac{a}{\left<a\right>} \right)$ is found ($\left<a\right>$ represents the average radius), independently of $\gamma$. All distributions in this range collapse then onto a single master curve as shown in (Fig. \ref{fig:steady_states}(b)). This behavior, along with the long tail in the distribution towards the right hand side, is typically observed in shear-fragmentation experiments in stirring tanks \cite{Spicer}. 

We note that our findings are robust with respect to the number of new particles formed by fragmentation. For instance, in Fig. 4(a) we see that the distributions of particles for ternary fragmentation are similar to the ones for binary splitting and only show a slight shift towards smaller size classes. The same result is found in population balance equation models, e.g. \cite{Spicer}.

Finally we mention that in spite of the different steady states, the size distribution in the initial phase is similar in the different cases. After short times, we find a roughly exponential decay. In this early phase, fragmentation is yet inactive, and the process is dominated by aggregation. This decay in the short time distribution can be found for all initial conditions.

In conclusion, we illustrated that an individual modeling of particles is able to reflect typical properties of aggregation/fragmentation processes. We found the development of a balance between aggregation and fragmentation, and a steady state. The steady-state particle size distributions found here correspond to those observed in rain drops (size-limiting fragmentation) and stirring tank experiments (shear fragmentation). For shear fragmentation the size distributions are found to follow a scaled form. 
In addition the approach shown here can reflect spatial inhomogeneity and take actual particle dynamics into account, and could possibly allow for a much more detailed description of particle interaction. It is thus more adequate than the usual stochastic, mean field like approach which relies on the assumption that the particles are well mixed \cite{prup}. The presence of chaotic attractors can ensure a partial mixing only and hence the assumption is not valid. 

An interesting open problem is to extend our study to 3D flows.

%\begin{acknowledgments}
The authors thank J. Br\"ocker, I. Geresdi, C. Grebogi, \'A. Horv\'ath and J. M\"arz for useful discussions and suggestions. The support of the Hungarian Science Foundation (OTKA T047233) is acknowledged. 
%\end{acknowledgments}


\begin{thebibliography}{99}
\bibitem{concentration}
For the cases studied here,
typical distances between particles are $\geq50a_1$.
\bibitem{obs}
For raindrops, typical numbers are $R \approx 10^{-3}$ and $a\approx 1$mm. For advection of these drops in a convective cloud cell of linear size $L\approx100$m and typical velocity fluctuations $U\approx1$m/s, we obtain $St \approx 2 \times 10^{-4}$ and $A \approx 5$, since $\nu \approx 10^{-5}m^2/s$ for air.
\end{thebibliography}
\end{document}